\documentstyle[12pt,aasms4]{article}

\def\etal{{\it et~al.\ }}
\def\kms{km~s$^{-1}$}

\begin{document}

\title{Redshift $z \sim 1$ Field Galaxies Observed
with the Keck Telescope and the HST \altaffilmark{1,2} }

\author{David C. Koo, Nicole P. Vogt, Andrew C. Phillips, Rafael Guzm\'an,
K.~L. Wu,  S.~M. Faber, Caryl Gronwall, Duncan A. Forbes, Garth D. Illingworth}
\affil{University of California Observatories / Lick Observatory, Board of
Studies \\ in Astronomy and Astrophysics, University of California, Santa
Cruz, CA 95064}

\author{Edward J. Groth}
\affil{Department of Physics, Princeton University, Princeton, NJ 08544}

\author{Marc Davis}
\affil{Department of Astronomy, University of California, Berkeley, CA
94720}

\author{Richard G. Kron}
\affil{Fermi National Accelerator Laboratory, MS 127, Box 500, IL 60510}

\and 

\author{Alexander S. Szalay}
\affil{Department of Physics and Astronomy, The Johns Hopkins University,
Baltimore, MD 21218}

\altaffiltext{1}{Based on observations obtained at the W. M. Keck Observatory,
which is operated jointly by the California Institute of Technology and the
University of California.}

\altaffiltext{2}{Based on observations with the NASA/ESA {\it Hubble Space
Telescope}, obtained at the Space Telescope Science Institute, which is
operated by AURA, Inc., under NASA contract NAS 5-26555}

\begin{abstract}

We report results based on 35 new spectroscopic redshifts obtained with the
Keck Telescope for field galaxies that also have photometry and morphology
from survey images taken by the refurbished Hubble Space Telescope.  A sample
of 24 redshifts for galaxies fainter than {\it I} $= 22$ has a median redshift
of $z \sim 0.81$.  This result is inconsistent with the lower median redshift
of $z \sim 0.6$ predicted by the ``maximal merger models'' of Carlberg (1996),
which otherwise fit existing data.  The data match an extrapolation of the
Canada France Redshift Survey (CFRS), as well as  predictions of certain mild
luminosity-evolution models.  Nearly half of the redshifts lie in two
structures at $z \simeq 0.81$ and $z \simeq 1.0$, showing the presence of high
density concentrations spanning scales of $\sim 1 h^{-1}$ Mpc, i.e., the size
of groups.  We find emission lines or the presence of possible  neighbors in 7
of 9 otherwise luminous galaxies with red central regions at redshifts beyond
$z \sim 0.7$. We also note a diversity of morphological types among blue
galaxies at $z \sim 1$, including small compact galaxies, ``chains,'' and
``blue nucleated galaxies.'' These morphologies are found among local, but
generally less luminous, galaxies. Distant blue galaxies also include
apparently normal late-type spirals.  These findings could imply modest bursts
of star formation caused by mergers or interactions of small, gas-rich
galaxies with each other or with larger, well-formed galaxies. This first
glimpse of very faint $z \sim 1$ field galaxies of diverse colors and
morphologies suggests that a mixture of physical processes is at work in the
formation and evolution of faint field galaxies.

\end{abstract}


\section{Introduction}

To date, we have yet to understand the origin of the large density of very
faint, blue field galaxies despite enormous observational progress (e.g.,
\cite{Lil95a}).  Deep  images to $I \sim 24$ or fainter with the refurbished
Hubble Space Telescope (HST)  show a predominance of late-type or unusual
galaxy morphologies (e.g., \cite{Gri94,For94,Gla95b,Dri95}), as well as small
sizes that suggest a large fraction of low-luminosity dwarf galaxies
(\cite{Dri95}).  The  redshifts of individual  galaxies observed with HST
remain largely unknown.  Existing redshift samples include 17 galaxies to $I
\leq 22$ from Forbes \etal (1996), 32 to $I \leq 22$ from Schade \etal (1995),
and 34 to $B \leq 24.5$ (roughly $I \leq 22.5$) from Cowie \etal (1995) but do
not yet probe fainter than  $I \sim 22$.  The redshift distribution is,
however, well established statistically to $I \leq 22$ from the CFRS (Lilly
\etal 1995b).  Redshifts are crucial for determining the intrinsic properties
of galaxies visible in deep HST images, and good spectra can also yield
rotation velocities, velocity dispersions, stellar population age indices, and
metallicities.  A program to obtain deep spectra using the 10-m Keck Telescope
is now underway as a new initiative called the Deep Extragalactic Evolutionary
Probe, or DEEP (\cite{Mou93,Koo95a}).

This paper reports on initial results of a new DEEP survey that utilizes
redshifts from the Keck Telescope, supplemented by photometry, colors, and
morphologies from  images taken by Groth \etal (1996) with the refurbished
HST.  The target sample contains 230 galaxies, but poor weather permitted only
18\% to be observed during the run.  The acquired redshift sample of 35
galaxies is, however, generally representative of the target sample. The
magnitudes are so faint (median $I \gtrsim 22$; median $B \sim 25$) and the
redshifts so high (median $z \sim 0.8$), that the current data provide a first
glimpse of the nature of faint, distant field galaxies of $\sim L^*$
luminosities at an epoch of roughly half the Hubble age.  \footnote{We adopt
$h = 0.75$, $q_o = 0$, and $\Lambda = 0$, i.e. $\Omega = 0$. At redshift $z
\sim 1$, $I \sim 22.6$ for a blue ($B-V \sim 0.65$) galaxy of $M_B \sim -20.4$
($M^*$), the lookback time is about 6.5 Gyr for a Hubble age of 13 Gyr, and 1
arcsec corresponds to 7 kpc.}

\section{Observations}

The HST data are from two surveys which we dub the ``Survey Strip'' taken
under HST program GTO 5090 (E. Groth as P.I.) and the ``Deep Field,'' under
GTO 5109 (J. Westphal as P.I.).  The Survey Strip consists of 28 overlapping
subfields taken with the HST Wide Field and Planetary Camera (WFPC2) and forms
a ``chevron strip'' oriented NE to SW at roughly 1417+52 at Galactic latitude
$b \sim 60$\arcdeg.  Each of  27 subfields has exposures of 2800s in the broad
$V$ filter (F606W) and 4400s in the broad $I$ filter (F814W). The 28th field
is the Deep Field (J2000 1417.5 + 52.5), with total exposures of 24400s in $V$
and 25200s in $I$. Astrometry and photometry of the entire Survey Strip has
been completed, with full details of the survey, data reductions, and
calibrations described by Groth \etal (1996). For the purposes of this paper,
magnitudes are ``total,''  while ($V-I$) colors are based on 1 arcsec diameter
apertures.

The morphologies were visually classified into three simple groups: ``O''
group for objects that appear round or elliptical, centrally concentrated,
smooth, and {\it largely} symmetrical with at most a very weak disk; ``S''
group for objects that have both a bulge and an apparent disk and/or spiral
arms; and ``*'' group for galaxies that are likely to have on-going star
formation with late-type, irregular, asymmetrical, multi-component, or
peculiar forms.  These group designations were chosen to match the symbols
used in the figures.  Note that at redshifts $z \sim 0.8$, the $I$ band
corresponds to rest-frame $B$.  Images of the entire spectroscopic sample are
shown in Figures~\ref{red}--\ref{misc}, along with  magnitudes, colors,
morphological group, redshifts (when known), and resulting absolute rest-frame
$B$ magnitudes and rest-frame $(U - B)$ colors. Table 1 provides J2000
positions and corresponding identifications from CFRS.

Except for two bright ($I \sim 18$) galaxies taken through clouds with a 1.0
arcsec-wide long slit, the other 39 galaxy targets and several setup stars
were observed through two focal-plane masks cut with multiple slitlets. This
multislit mode allows simultaneous exposure of $\sim 25$ or more targets  with
the Low Resolution Imaging Spectrograph (LRIS; see \cite{Oke95}).  The
detector was a Tektronix 2048$\times$2048 px CCD (24 $\micron$ or 0.215 arcsec
px$^{-1}$). With a slitwidth of 1.23  arcsecs, a 600 l/mm grating yielded 1.26
\AA\ px$^{-1}$ and an instrumental resolution of $\sim 4$ \AA\ FWHM, i.e., R
$\sim$ 2000.  The spectral range was about 6500 \AA\ -- 9100 \AA, depending on
the exact position of the target on the mask.  This spectral range was chosen
to detect the strongest emission lines in faint galaxies: H${\alpha}$
$\lambda$6563 at $z \la 0.4$, H$\beta$ $\lambda$4861 and [OIII] $\lambda$5007
at $0.3 \la z \la 0.8$, and [OII] $\lambda$3727 at $0.75 \la z \la 1.4$.  The
slitlets were aligned along the 7 arcmin long axis of the LRIS field, which in
turn was aligned with the NE-SW orientation of the Survey Strip. Of the four
allotted nights, only part of one night, UT 1 May 1995, was clear and calm
enough to open the dome, during which we acquired 4$\times$1800s exposures
through one mask and 3$\times$1800s through another at the same position (but
with a different selection of galaxies).  Seeing was variable but was
estimated to be about 0\farcs 8 to 1\farcs 2  FWHM during most of the
exposures.  The spectroscopic reduction included the usual corrections for
bias, dark, flats, and cosmic rays as well as sky-line wavelength calibrations
and background sky subtraction.  We chose to place the targets at two
positions along the slit, separated by 3 arcsecs, during different exposures.
This ``dither'' allowed corrections for the 1-2\% fringing of the night sky
seen at wavelengths greater than $\sim$ 7200\AA\, by subtraction of the
dithered images taken at each position.  This method increases random sky
noise by a factor of ~$\sqrt[]{2}$.  

The 230 galaxies chosen for spectroscopy, of which only 38 were actually
observed with Keck, do not constitute a totally random, magnitude-limited
sample. Rather, they were chosen to span the full distribution of colors and
diverse visual morphologies of the galaxies in the whole Survey Strip. The
field of our Keck observations included the Deep Field, where faint galaxies
($24 \le (V+I)/2 < 25$) were selected, and four adjacent WFPC2 subfields,
where brighter targets ($(V+I)/2 < 24$) were selected (see Figure~4).  As
indicated in Table 1, a few special targets were chosen because of their
unusual morphology (e.g., a promising quadruple-lens candidate); their
detection in the radio; or their elongation with position angles promising for
rotation curve measures (Vogt \etal 1996).  In addition, redshifts were
measured for four emission-line objects serendipitously lying within slitlets
of intended targets.  The faintest objects were observed through both masks. 
Within small number statistics and as shown in Figure~4, the present Keck
sample  has a color distribution at each magnitude roughly similar to the
entire HST sample, except that ten galaxies (rather than the expected seven)
were observed with  very red colors ($V - I > 1.6$) and magnitudes $20 < I <
23.25$.

The redshift identifications were made by first visually inspecting the
two-dimensional sky-subtracted images for any strong emission lines. The
extracted one-dimensional spectra were then examined for absorption lines when
the continuum was bright enough, generally for galaxies with $I \lesssim 23$,
or weaker emission features. In total, we have extracted 24 redshifts for our
original targets, four redshifts for objects fortuitously lying in a slitlet,
and two redshifts from longslit observations of $I \sim 18$ galaxies.  The
reliability of these redshifts range from high (i.e., secure redshifts) to
moderately high (i.e., probable redshifts).  In several galaxies, only one
emission line was detected, but its identification as [OII] $\lambda$3727 was
considered to be secure if we resolved the doublet or found  a rising
continuum to the red.  Five more objects have apparent emission or absorption
features, but highly uncertain redshifts; six targets fainter than $I \sim
22.3$ have yet to yield any reliable spectral features for redshift
determination.  Figure 5 shows examples of our Keck spectra that span a range
of redshift reliability.  
  
An independent check of our coordinates, photometry, and redshifts can be made
from the eight galaxies in common with CFRS (see Table 1 and Lilly \etal
1995a) and shows good agreement.  We find a mean positional offset between the
CFRS and our coordinates of $\sim 1\farcs 1 \pm 0\farcs4$, which is certainly
close enough to make secure identifications. The mean $I$ band difference
(CFRS $I_{AB} - I_{tot}$) is expected to be $\sim 0.48$ mag, while we find
0.56 $\pm$ 0.33; dropping the two very bright, large galaxies, 074-2237 and
074-2262, whose photometry might be difficult, yields a significantly smaller
standard deviation ($\pm$ 0.18) and a mean within expectations (0.46 mag).
Finally, the redshifts of all eight CFRS galaxies in common with our Keck
sample agree well individually,  but have slightly lower values than ours by a
mean difference of $\delta z = 0.0021 \pm 0.0025$ (see Table 1).

\section{Results}

The Keck redshift survey is 100\% complete for targeted objects to $I \sim
22.3$ (24 galaxies) and 85\% complete for the whole sample to $I \lesssim
24.5$. We reached beyond $I \sim 23$ only for targets with strong emission
lines; the redshifts of several of these faint galaxies remain highly
uncertain (see Table 1).  Figure~6b shows the magnitude versus redshift
distribution of the sample, plus the 17 galaxies from Forbes \etal (1996).

The most striking aspect of the data is the high fraction of redshifts $z
\gtrsim 0.8$.  The resultant median of the observed redshifts is 0.81.  Due to
the relatively high level of completeness, the median value of the whole
sample must lie in the range 0.81 to 1.00, even if the undetermined redshifts
are entirely at $z < 0.6$ or at  $z > 1.4$, or if the number of galaxies at
the $z = 0.81$ and 1.00 peaks  are halved to account for small-scale
clustering effects.  

For comparison, the  CFRS sample (\cite{Lil95a}) has nearly 600 redshifts to
$I = 22$, a median redshift of $z \sim 0.56$, an overall completeness of 85\%,
and completeness for $21 < I < 22$ of 74\%.  Our sample is 100\% complete to
these limits and includes 25 redshifts beyond $z \sim 0.8$, more than the 4
such redshifts in the $B \le 24$ survey of Glazebrook \etal (1995a), the 13 in
the $K \leq 20$ survey of Songaila \etal (1994), or the 5 in the $I \le 22$
targets of Schade \etal (1995), but still far short of the $\gtrsim150$ in the
CFRS.  

Figure~7 plots ($V-I$) color versus redshift for the Keck sample with
morphological information encoded within the symbol shapes. The curves provide
a guide to the intrinsic colors of local galaxies and to the expected colors
for single instantaneous starbursts originating at redshifts $z = 1$ or $z =
2$.  One striking aspect of this figure is how well the galaxies fall within
the bounds seen in local galaxies, a result already found in the larger CFRS
sample (Crampton \etal 1995).  We find no evidence for a significant new
population of unusually blue or unusually red galaxies.  Other straightforward
results are that very red {\it field} galaxies do exist at high redshifts $z
\ga 0.7$ and that such galaxies have colors comparable to those found for
local ellipticals; therefore, their most recent major star-formation event
presumably occurred at redshifts $z \ga 2$.  Moreover, the intrinsically
reddest galaxies are not necessarily purely elliptical in morphology; bulges
and at least one dust-reddened, inclined spiral are seen.  Although
low-clustering amplitudes have been measured in angular correlation studies
(e.g., Efstathiou \etal 1991), the intrinsically blue galaxies here do partake
in the strong clustering seen at redshifts $z \sim 0.8$ and $1.0$. They may
thus contribute to the strong angular clustering observed among the bluest
galaxies by Landy \etal (1996). The diverse morphologies of these
high-redshift, clustered galaxies are easily seen in Figures~1 and 2.


\section{Discussion}

\subsection{Redshift Distribution}

What cosmological implications can we extract from this  faint sample of
galaxies? Given the relatively small numbers, median redshifts are a good
starting point for comparison to theory. First, the 100\% complete sample of 9
galaxies  with $19.5 < I < 22$ has a median redshift of $z = 0.81$, with a
possible range from 0.37 to 1.00 at the 95\% confidence level; this is
consistent with the median of 0.56 for CFRS based on a much larger sample. 
Second, the $I \ge 22$ sample with 24 galaxies has the same median of 0.81 but
a tighter range of $z \sim 0.81$ to 1.00 at the 95\% confidence level. This
high median is an important finding, since Songaila \etal (1994) found the
median to remain at $z \lesssim 0.6$ for galaxies fainter than $K \sim 18$
(roughly $I \sim 20.5$), as predicted by some merger models.  For example, the
``maximal merger model'' of Carlberg (1996), which matches observations to $I
\sim 22$, predicts that the median will stabilize at $z \sim 0.6$ from $I \sim
20$ all the way to $I \sim 28$, contrary to what we find in this sample at $I
> 22$.  The high median is, however, consistent with at least two other
scenarios.  First, with no further evolution than that already found in the
CFRS survey, Lilly \etal (1995c) predict that the median should continue to
rise to $z \sim 0.9$ for $22 < I < 23$ and to $z \gtrsim 1.0$ for $23 < I <
24$. Second, the predictions of models that do include mild luminosity
evolution but no mergers, such as those of Gronwall \& Koo (1995), are
virtually identical  to those predicted by CFRS.  At this time, we have no
basis to exclude other models, such as those of Cole \etal (1994), that
analytically track various other physical processes such as supernova gas
removal and star-formation, as well as merging of dark matter halos.  

The width of the distribution is also a potentially powerful discriminant. For
example, the fiducial bursting dwarf model of Babul \& Ferguson (1996)
predicts a narrow redshift distribution peaked at $z \la 1$ for a $B < 26$
redshift sample, with a cutoff in numbers at $z = 1$, chosen by them by fiat,
when dwarfs first form.  Our data, with the bulk of the redshifts $0.7 \la z
\la 1$, lend some support for this picture.  On the other hand, several very
blue galaxies are seen beyond $z = 1$, while those very near $z = 1$ span a
wide range of colors rather than being dominated by the very blue colors
expected from an ongoing burst of star formation (see Figure 7).  Moreover,
the diverse morphologies and diffuseness of many of the blue galaxies,
discussed below, may be difficult to explain.  More faint redshifts with
stricter selection criteria will be needed to provide more definitive tests of
the bursting model.  

Redshifts also probe clustering. CFRS already paved the way with the discovery
of 12 redshifts in a 0.016 redshift interval, indicating a five to ten times
overdensity of galaxies at $z = 0.985$ (Le F\`evre \etal 1994).  This possible
supercluster structure was found in a 10'x10' field that coincidentally 
overlaps our field.  As seen in Figure~6a, we find a density enhancement at $z
= 0.995 \pm 0.004$ containing five galaxies, which may be the same structure,
plus another stronger one at $z = 0.811 \pm 0.003$ with ten galaxies (standard
deviations are measured within $\delta z = \pm 0.02$ of the peak).  Since we
see no visual evidence for any rich cluster of galaxies in our field and the
velocity dispersion is relatively low ($\sim 450$~\kms), the $z = 0.81$
enhancement is probably a rich group.

\subsection {Morphologies, Colors, and Galaxy Evolution}

We close with some discussion and speculations based on  HST morphologies and
colors.  Figure~1 shows 9 galaxies with very red colors within their central
one arcsec diameters.  The galaxies in the top two rows of Figure~1 appear
morphologically to be elliptical or S0 galaxies (though Phillips \etal [1996]
find that 073-2675 is better fit with an exponential, rather than an $r^{1/4}$
profile).  Combined with very red colors, these galaxies indicate the presence
of luminous ellipticals {in the field} that are already quite old by $z \sim
1$.  This presumes that the red colors are not due to dust extinction, which
remains to be determined.  Based on the $z = 2$ curve in Figure~7, the last
significant star formation event presumably occurred at redshifts $z \sim 2$
or greater.  Such very red galaxies have also been found among distant radio
galaxies (e.g., McCarthy 1993) and cluster galaxies (e.g., Dickinson 1996).

Yet several of these galaxies are not quite ``normal.'' Galaxy 093-2470
appears to have four very blue compact objects imbedded in a symmetrical
pattern within its halo, a pattern proposed to form a ``quad-lens'' system
(Ratnatunga \etal 1995, Broadhurst \etal 1996, Crampton \etal 1996).  Although
the ``maximal merger model'' of Carlberg (1996) appears to be invalidated by
the new redshifts, we do find some evidence for minor  mergers among several
of these distant red galaxies.  Except for 073-2675, the remaining five
galaxies in the first two rows of Figure~1 show close projected neighbors or
tidal and merger ``tails,'' and some of this neighboring material is blue.
Perhaps we are watching the infall of dwarf galaxies, some possibly quite
gas-rich, if the blue colors are due to active star formation.  The
implication would be that elliptical galaxies form early and yet can be built
up by minor mergers over a much longer time period.  This scenario could
reconcile  the apparent lack of luminosity or density evolution among red
galaxies seen in the CFRS (Lilly \etal 1995c) or the Mg II absorber sample
(Steidel, Dickinson, \& Persson 1994) with the presence of well-formed, red
field galaxies at an early epoch.  

The last row of Figure~1 presents three very red {\it disk} galaxies.  With a
peak-to-peak rotation curve velocity of $\sim$600~\kms ~(Vogt \etal 1996), the
conspicuous edge-on system (104-4024) demonstrates that some massive,
thin-disk, dusty systems already exist at $z \sim 0.8$, roughly half a Hubble
age ago.  This single object, evidently similar to our Milky Way, supports 
what we believe to be the early formation of our own old disk. There are hints
of very faint blue satellites that might eventually settle into this distant
spiral, an evolutionary path that has also been proposed for the Milky Way
(Majewski 1993).

Another disk-like system, 094-2210, shows a very red, bulge-like core
surrounded by numerous blue ``blobs'' or arms.  If this galaxy is a disk or
proto-disk system with little dust extinction, it suggests that some bulges
are already quite old by $z \sim 1$, with their most recent star formation
occurring perhaps at $z \ga 2$ (see Figure~7).  Object 103-2074 appears to be
another disky system of early type (S0 or Sa) with a very red bulge.

Figure~2 shows two sets of galaxies: the bluer, high-redshift galaxies and
those without redshifts. Those with measured redshifts $z \gtrsim 0.75$ are
organized into 7 rows divided into morphological groups.  We argue that all
six of the galaxies without redshifts are likely to be at high redshifts $z
\ge 0.75$.  Since blue galaxies normally have strong emission lines of [OII]
$\lambda$ 3727 that should be detectable to $z \approx 1.4$ (or especially of
H${\alpha}$ for redshifts lower than $z \sim 0.4$), we argue that the five
blue galaxies in our Keck sample without measured redshifts are probably at
higher redshifts.  In support of this conjecture, we note that the bluest
galaxy in our entire Keck sample (source 084-1720) has two definite absorption
lines separated by $\sim$35\AA\ and appearing around 6770\AA\ (see Figure 5);
we tentatively identify this doublet as Fe II $\lambda$2587 and $\lambda$2600
that would yield a redshift of $z = 1.60$.  One relatively bright, very red
galaxy (source 083-3138) also has no redshift, but based on the similarity of
its color to the galaxies in Figure 1 with redshifts (see Figure 7), we infer
that it is likely to be at high redshift.  The targets without redshifts are
all presented in the last two rows of Figure~2.

The large diversity of morphologies displayed in Figure~2 is striking, ranging
from compact galaxies shown in the first and second rows, normal disk-like
systems in the third and fourth rows, to irregular, late-type systems with
multiple blue star-formation sites in the fifth and sixth rows. This last
group could  be related to a proposed new class of ``blue-nucleated'' galaxies
(Schade \etal 1995).  The seventh row shows galaxies that are perhaps more
elongated versions of the above galaxies or that may belong to the new class
of ``chain'' galaxies defined by Cowie \etal (1995).  Comparably complex and
diverse morphologies can, however,  be found even among local very blue
galaxies such as clumpy irregulars, Markarian galaxies, extragalactic HII
regions, merging galaxies, or other peculiar classes (c.f. images of HII
galaxies presented by Melnick 1987).  These various local blue systems often
show similar, multiple concentrations of intense star formation.  

The new Keck redshifts and HST morphologies combine to  counter the  view that
low redshift, very blue, low luminosity dwarf galaxies dominate the late-type,
peculiar systems detected in deep HST images (e.g., Driver \etal 1995). 
Instead, most of the Keck galaxies with these morphologies are at high
redshifts and are thus relatively luminous, though typically less luminous
than $L^*$.  The apparent sizes of many of these distant galaxies are $\sim 1$
arcsec, however,  which imply metric sizes of several kiloparsecs, quite
typical of local dwarf galaxies.  After corrections for cosmological dimming
(about 2.3 magnitudes at $z \sim 1$) and K-corrections (from $\sim$ 0 mag for
very blue galaxies  to over 4 mag for very red galaxies at $z \sim 1$), the
resultant restframe $B$ band surface brightnesses are higher than those of
most local spirals and star-forming irregulars (Schade \etal 1995, Phillips
\etal 1996).  Thus, the faint blue  galaxies  are not explained by a class of
very low surface brightness galaxies that have been missed by local surveys
(McGaugh 1994).  Whether the high redshift  galaxies are indeed low-mass
systems (which  may or may not have any direct relationship to local dwarfs)
or are the luminous tips of more massive systems will be decided by results of
kinematic surveys.  So far, only small samples of very blue, distant, field
galaxies have been observed, many quite compact and as distant as $z \sim
0.8$, and these have yielded some luminous galaxies with very-low,
emission-line velocity widths of $\sigma \la 70$~\kms (Koo \etal 1995, Colless
1995, Guzm\'an \etal 1996, Forbes \etal 1996).

We conclude from the present sample that no single physical process (e.g.,
mergers or bursting dwarfs) dominates the evolution of faint galaxies at
redshifts $z \sim 1$.  We find a diversity of morphologies from normal to
peculiar, with no firm evidence for entirely new classes of galaxies for which
local counterparts cannot be found, a conclusion also reached by Forbes \etal
(1996).  We also find bulge and elliptical systems that are well formed and
red.  Equally tantalizing are the hints that a large fraction of galaxies are
participating in further agglomeration and continued star-formation due to
interactions, major mergers, and infalling satellites. Taken together, these
early Keck results for distant HST galaxies imply that much larger samples
will be needed to disentangle and understand this exciting, complex, and very
important early history of galaxies.

\acknowledgements

We are grateful to the staff of the Keck Observatory, especially T. Bida, for
their assistance, as well as J. Lewis at the UCO/Lick Observatory Laboratories
for fabricating the masks.  The Center for Particle Astrophysics initiated the
DEEP program and continues to provide support. We also thank C. Steidel and R.
Blandford for their comments and participation in DEEP and Jason Rhodes, Rick
Balsano, and the WFPC-1 Investigation Definition Team for pre-publication
access to their catalog and HST images.  We are grateful to the referee for
several useful suggestions.  Funding for this work was provided by NSF grants
AST-922540, AST-9120005, and AST-8858203; NASA grants AR-5801.01-94A,
GO-2684.04-87A, GO-2684.05-87A, and WFPC-1 Investigation Definition Team Grant
Nos. NAS5-1661 and NAS5-1629. R. G. acknowledges funding from the Spanish MEC
fellowship EX93-27295297; C.G. from an NSF Graduate Fellowship.

\clearpage

\begin{table}
\dummytable\label{tbl-1}
\end{table}

\clearpage

\figcaption[] 
{Distant Keck spectroscopic targets with red colors ($V_{606} - I_{814} >
1.75$) within the central 1 arcsec diameter and with high redshifts, $z >
0.75$.  Each image ($V_{606}$ on left and $I_{814}$ on right) covers 8 x 9
arcsec (unless cut off by the edge of WFPC2) and is the result of Hanning
smoothing of the original.  Each image pair is labeled by (first line) source
identification number (see Table 1), total $I$ mag, visual morphology group
(0, S, or *) as discussed in the text, observed ($V - I$) color, then (second
line) redshift followed by absolute luminosity ($M_B$) and approximate
rest-frame ($U-B$) color.  Highly uncertain values are marked with a ``:''.
The cardinal directions are shown in the lower righthand corner, and the image
scale is marked in the upper lefthand corner.  
\label{red}
}

\figcaption[]
{Similar to Fig.~1, but showing high-redshift targets with blue colors
($V_{606} - I_{814} < 1.75$) or galaxies with no redshifts.  The first set of
blue, high-redshift galaxies (in rows 1 - 7) is grouped roughly by similar
morphology as follows: row 1 -- very compact galaxies; row 2 -- slightly more
diffuse but still compact; row 3 -- spirals; row 4 -- possible spirals; rows 5
and 6 -- diffuse galaxies with multiple knots of star-formation that might be
similar to the ``blue nucleated galaxy'' class proposed by Schade \etal
(1995); row 7 -- elongated diffuse galaxies that are similar to the ``chain''
class identified by Cowie \etal (1995).  The second set of six targets (in
rows 8 and 9)  have no redshifts. As discussed in the text, we believe the one
very red, relatively bright galaxy and five blue galaxies all belong to the 
high-redshift group.
\label{blue} 
}

\figcaption[]
{Similar to Figs.~1 and 2, but  for lower redshift galaxies ($z < 0.75$).  The
images of the two galaxies at a redshift lower than 0.2 are 12.5 x 15 arcsec,
and those of the five galaxies with redshifts between 0.2 and 0.75 are 5 x 9
arcsec.  
\label{misc} 
}

\figcaption[] 
{($V-I$) vs. $I$ plot showing a complete sample of objects found within fields
4 to 10 of the Survey Strip (dots); those chosen as the original Keck targets
(circled dots); those for which Keck spectra and redshifts were actually
acquired (solid circle); and those with Keck spectra but no redshifts
(triangles).  The two lines ($(V + I)/2 = 24$, $(V + I)/2 = 25$) mark the
primary selection criteria for objects in the Deep Field; brighter targets
were selected from the surrounding fields with the constraint $(V + I)/2 < 24$.
\label{VIvI} 
}

\figcaption[]
{Counts vs. wavelength for sample Keck spectra, each labeled with the source
name, total $I_{814}$ mag, and redshift.  The top one is for an 1800s exposure
of the night sky.  The next two are galaxies whose redshifts are secure, with
one being a bright galaxy with several strong absorption lines and the other
being a 2.6 mag fainter galaxy with resolved [OII] emission lines and [OIII].
The next galaxy is another magnitude fainter with an emission line that we
identify as [OII], but since it has no other signatures to confirm this
identification, its redshift reliability is only probable.  The last galaxy is
bright enough to yield definite absorption lines, but we feel the
identification of these with Fe II is tentative, and thus highly uncertain, at
this time.  The strengths of the spectral features in these galaxy spectra, 
all with two-hour exposures and smoothed with a seven-pixel boxcar (except for
084-1720 which was smoothed with a fifteen-pixel boxcar), are higher than
average.
\label{spectra}
}

\figcaption[]
{Upper panel (a) shows the redshift histogram for the Keck targets from this
work  in bins of 0.02 in $z$, while the lower panel (b) shows an $I$ magnitude
vs. redshift z (on a  logarithmic scale)  plot from this work, plus 17
galaxies from Forbes \etal (1996).  For comparison to models, the solid line
in (b) shows the median curve predicted from the evolving model of Gronwall \&
Koo (1995); the dashed line shows the median curve predicted from a standard
no-evolution model (Guiderdoni \& Rocca-Volmerange 1990). The dotted line
shows the predicted median curve from the ``maximal merger model'' of Carlberg
(1996). For discriminating the relative luminosities of the galaxies, the
dash-dot lines have been included to show the track for an $M^*$ and $M^*+2$
galaxy that has the color of an Sbc galaxy.  Objects without redshifts are
placed in the box at their respective magnitude positions.  
\label{zvsI}
}

\figcaption[]
{($V-I$) color vs. redshift for  Keck targets from this work.  Objects without
redshifts are placed in the box to the right.  Morphologies are indicated by
symbols ``O'', ``S'', and ``*'' that match the visual morphology groups
discussed in the text.  Several labeled lines show expected colors for various
spectral energy distributions: a) an instantaneous burst of star formation at
redshift $z \sim 2$ (using the models of Bruzual and Charlot 1993) that
becomes almost as red as b) a  {\it non-evolving} local elliptical or S0
(E/S0) by $z < 1$; c) model burst at $z = 1$ that might be compared to the
bursting dwarfs in the fiducial model of Babul and Ferguson (1996); d) a
non-evolving local Sbc galaxy; and e) the bluest one, derived from a
non-evolving  spectrum of NGC 4449, a nearby, very actively star-forming Sm
galaxy.  
\label{VIvsz}
}


\clearpage

\begin{thebibliography}{}

\bibitem[Babul \& Ferguson 1996] {Bab96} Babul, A., \& Ferguson, H. C. 1996,
\apj, 458, 100

\bibitem[Broadhurst \etal 1996] {Bro96} Broadhurst, T. J., \etal 1996, in
preparation

\bibitem[Bruzual \& Charlot 1993] {Bru93} Bruzual, G., \& Charlot, S. 1993,
\apj, 405, 538

\bibitem[Carlberg 1996] {Car96} Carlberg, R. G. 1996, in {\it Galaxies in the
Young Universe} MPI Conference, ed. H. Hippelein (Springer), preprint

\bibitem[Cole \etal 1994] {Col94} Cole, S., Aragon-Salamanca, A., Frenk, C.
S., Navarro, J. F., \& Zepf, S. E. 1994, \mnras, 271, 781

\bibitem[Colless 1995] {Col95} Colless, M. 1995, in ``Wide Field Spectroscopy
and the Distant Universe,'' eds. S. Maddox and A. Arag\'on-Salamanca (World
Scientific: Singapore), 263

\bibitem[Cowie \etal 1995] {Cow91} Cowie, L. L., Hu, E. M., \& Songaila, A.
1995, \aj, 110, 1526

\bibitem[Crampton \etal 1995] {cram95} Crampton, D., Le F\`evre, O., Lilly, S.
J., \& Hammer, F. 1995, \apj, 455, 96

\bibitem[Crampton \etal 1996] {cram96} Crampton, D., Le F\`evre, O., Hammer,
F., \& Lilly, S. J. 1996, \aap, preprint

\bibitem[Dickinson 1996] {Dic96} Dickinson, M. 1996, in ``Fresh Views on
Elliptical Galaxies,'' eds. A. Buzzoni, A. Renzini, and A. Serrano (A.S.P.
Conf. Ser. 86), 283

\bibitem[Driver \etal 1995] {Dri95} Driver, S. P., Windhorst, R. A.,
Ostrander, E. J., Keel, W. C., Griffiths, R. E., \& Ratnatunga, K. U. 1995,
\apj, 449, L23

\bibitem[Efstathiou \etal 1991] {Efs91} Efstathiou, G. P. E., Bernstein, G.,
Katz, N., Tyson, J. A., \& Guhathakurta, P. 1991, \apj, 380, L47

\bibitem[Fomalont \etal 1991] {Fom91} Fomalont, E. B., Windhorst, R. A.,
Kristian, J. A., and Kellerman, K. I. 1991, \aj, 102, 1258

\bibitem[Forbes \etal 1994] {For94} Forbes, D. A., Elson, R. A. W., Phillips,
A. C., Koo, D. C., \& Illingworth, G. D. 1994, \apj, 437, L17

\bibitem[Forbes \etal 1996] {For96} Forbes, D. A., Phillips, A. C., Koo, D.
C., \& Illingworth, G. D. 1996, \apj, in press

\bibitem[Glazebrook \etal 1995a] {Gla95a} Glazebrook, K., Ellis, R., Colless,
M., Broadhurst, T., Allington-Smith, J., \& Tanvir, N. 1995a, \mnras, 273, 157

\bibitem[Glazebrook \etal 1995b] {Gla95b} Glazebrook, K., Ellis, R., Santiago,
B., \& Griffiths, R. 1995b, \mnras, 275, L19

\bibitem[Griffiths \etal 1994] {Gri94} Griffiths, R. E. \etal 1994, \apj, 437,
67

\bibitem[Gronwall \& Koo 1995] {Gron95} Gronwall, C., \& Koo, D. C. 1995,
\apjlett, 440, L1

\bibitem[Groth \etal 1996] {Gro96} Groth, E. J., \etal 1996, in preparation

\bibitem[Guzman \etal 1996] {Guz96} Guzm\'an, R., Koo, D. C., Faber, S. M.,
Illingworth, G. D., Takamiya, M., Kron, R. G., \& Bershady, M. A.  1996, \apj,
in press

\bibitem[Guiderdoni \& Rocca-Volmerange 1990] {Gui90} Guiderdoni, B., \&
Rocca-Volmerange, B. 1990, \aap, 227, 362 

\bibitem[Koo 1995] {Koo95a} Koo, D. C. 1995, in ``Wide Field Spectroscopy and
the Distant Universe,'' eds. S. Maddox and A. Arag\'on-Salamanca (World
Scientific: Singapore), 55

\bibitem[Koo 1995b] {Koo95b} Koo, D. C., Guzm\'an, R., Faber, S. M.,
Illingworth, G. D., Bershady, M. A., Kron, R. G., \& Takamiya, M. 1995, \apj,
440, L49

\bibitem[Landy \etal 1996] {Lan96} Landy, S. D., Szalay, A. S., \& Koo, D. C.
1996, \apj, in press

\bibitem[Le F\`evre \etal 1994] {LeF94} Le F\`evre, O., Crampton, D., Hammer,
F., Lilly, S. J., \& Tresse, L. 1994, \apj, 423, L89

\bibitem[Lilly \etal 1995a] {Lil95a} Lilly, S. J., Hammer, F., Le F\`evre, O.,
\& Crampton, D. 1995a, \apj, 455, 75

\bibitem[Lilly \etal 1995b] {Lil95b} Lilly, S. J., Le F\`evre, O., Crampton,
D., Hammer, F., \& Tresse, L. 1995b, \apj, 455, 50

\bibitem[Lilly \etal 1995c] {Lil95c} Lilly, S. J., Tresse, L., Hammer, F., Le
F\`evre, O., \& Crampton, D.  1995c, \apj, 455, 108

\bibitem[Majewski 1993] {Maj93} Majewski, S. R. 1993, \araa, 31, 575

\bibitem[McCarthy 1993] {McC93} McCarthy, P. J. 1993, \araa, 31, 639

\bibitem[McGaugh  1994] {McG94} McGaugh, S. S. 1994, Nature, 367, 538

\bibitem[Melnick  1987] {Mel87} Melnick, J. 1987, in ``Starbursts and Galaxy
Evolution,'' eds. T. X. Thuan, T. Montmerle, J. T. Thuan Van (Edition
Frontiers: Singapore), 215

\bibitem[Mould 1993] {Mou93} Mould, J. 1993, in ``Sky Surveys: Protostars to
Protogalaxies,'' ed. B. T. Soifer, (ASP Conf. Ser. 43), 281

\bibitem[Oke \etal 1995] {Oke95} Oke, J. B., \etal 1995, \pasp, 107, 375

\bibitem[Phillips \etal 1996] {Phi96} Phillips, A. C. \etal 1996, in
preparation

\bibitem[Ratnatunga \etal 1995] {} Ratnatunga, K. U., Ostrander, E. J.,
Griffiths, R. E., \& Im, M. 1995, \apj, 453, L5

\bibitem[Schade \etal 1995] {Sch95} Schade, D., Lilly, S. J., Crampton, D.,
Hammer, F., Le Fe\`vre, O., and Tresse, L. 1995, \apj, 451, L1

\bibitem[Songaila \etal 1994] {Son94} Songaila, A., Cowie, L. L., Hu, E. M.,
\& Gardner, J. P. 1994, \apjs, 94, 461

\bibitem[Steidel \etal 1994] {Ste94} Steidel, C. C., Dickinson, M., \&
Persson, S. E. 1994, \apj, 437, L75

\bibitem[Vogt \etal 1996] {Vog96} Vogt, N. P., Forbes, D. A., Phillips, A. C.,
Gronwall, C., Faber, S. M., Illingworth, G. D., \& Koo, D. C. 1996, \apjl,
accepted

\end{thebibliography}
\end{document}